\documentclass{article}

\usepackage{arxiv}

\usepackage[utf8]{inputenc} 
\usepackage[T1]{fontenc}    
\usepackage{hyperref}       
\usepackage{url}            
\usepackage{booktabs}       
\usepackage{amsfonts}       
\usepackage{nicefrac}       
\usepackage{microtype}      
\usepackage{lipsum}		
\usepackage{graphicx}
\usepackage{natbib}
\usepackage{doi}
\usepackage{amsmath}
\usepackage{natbib}

\title{Development of a Machine Learning based Radio source localisation algorithm for Tri-axial antenna configuration}

\author{Harsha Avinash Tanti, Abhirup Datta, Tiasha Biswas, Anshuman Tripathi\\
       Department of Astronomy, Astrophysics, and Space Engineering, Indian Institute of Technology Indore,\\ Madhya Pradesh, 453552, India.\\ \texttt{phd1901121009@iiti.ac.in}\\
}

\date{}

\renewcommand{\shorttitle}{\textit{arXiv} Template}

\hypersetup{
pdftitle={A template for the arxiv style},
pdfsubject={q-bio.NC, q-bio.QM},
pdfauthor={David S.~Hippocampus, Elias D.~Striatum},
pdfkeywords={First keyword, Second keyword, More},
}

\begin{document}
\maketitle

\begin{abstract}
	Accurately determining the origin of radio emissions is essential for numerous scientific experiments, particularly in radio astronomy.
    Conventional techniques, such as the use of antenna arrays encounter significant challenges, specially at very low frequencies, due to factors like the substantial size of the antennas and ionospheric interference.
    To address these challenges, we employ a space-based single-telescope that utilizes co-located antennas, complemented by goniopolarimetric techniques for precise source localization.
    This study explores a novel and elementary machine learning (ML) technique as a way to improve and estimate Direction of Arrival (DoA), leveraging a tri-axial antenna arrangement for radio source localization.
    Employing a simplistic emission and receiving antenna model, our study involves training an artificial neural network (ANN) using synthetic radio signals.
    These synthetic signals can originate from any location in the sky and cover an incoherent frequency range of 0.3 to 30 MHz, with a signal-to-noise ratio (SNR) between 0 and 60 dB.
    Then, a large data set was generated to train the ANN model catering to the possible signal configurations and variations.
    After training, the developed ANN model demonstrated exceptional performance, achieving loss levels in the training ($\sim0.02$), validation ($\sim0.23\%$), and testing ($\sim0.21\%$) phases.
    The machine learning-based approach remarkably, exhibits substantially quicker inference times ($\sim5$ ms) in contrast to analytically derived Direction of Arrival (DoA) methods, which typically range from 100 ms to a few seconds.
    This underscores its practicality for real-time applications in radio source localization, particularly in scenarios with limited number of sensors.
\end{abstract}


\section{Introduction}

The localization of radio sources, also called Direction of Arrival (DoA), refers to the direction of emission of electromagnetic (EM) sources or radio sources \citep{Cecconi2005, Tanti2023}.
Despite being a challenging task, the localization of EM waves is particularly crucial in radio astronomy.
It contributes significantly to the identification of emission characteristics and offers valuable information on radio sources \citep{Cecconi2005, Chen2010}.

Radio bands, ranging from around metres to millimetres waves, have been extensively studied using radio telescopes (such as FAST) and interferometer facilities (such as VLA, GMRT) \citep{Lacy_2020, Swarup1990, Qian2020}.
However, observing radio emissions at frequencies below 30 MHz, particularly under 16 MHz, is difficult due to the ionospheric cutoff, making these frequencies unexplored in radio astronomy \citep{Bentum2011}.
To overcome this limitation and conduct sensitive observations, space or moon-based radio telescopes are the most suitable alternatives \citep{Bentum2011}. 
Having a space-based array or interferometer is ideal for source localisation.
Single satellite systems such as Cassini Radio and Plasma Wave Science (RPWS), Solar TErrestrial RElations Observatory (STEREO), the Netherlands-China Low-Frequency Explorer (NCLE) and the Space Electric And Magnetic Sensor (SEAMS) are preferred owing to engineering challenges and financial constraints \citep{Weiler2000, Cecconi2005, Chen2010, Tanti2023}.
Antenna arrays and interferometers utilize methods such as triangulation, while a single radio telescope uses co-located antenna configurations and gonio-polarimetric methods for estimating DoA \citep{Chen2010}.
Space projects like Cassini-RPWS, STEREO/WAVES, and NCLE employ gonio-polarimetric methods for source localisation \citep{Cecconi2005, Chen2010}. 
For DoA estimation, the orientation of the sensors (antenna) is crucial \citep{Tanti2023}.

Antenna/Sensor array methods like MUltiple SIgnal Classification (MUSIC), Estimation of Signal Parameters via Rotational Invariance Technique (ESPRIT), and Root-MUSIC resolve DoA but with high computational complexity and inference time \citep{Waweru2014}. 
Conversely, Pseudo-vector, Analytical Inversion, and MPM-based DoA offer computationally efficient alternatives \citep{Carozzi2000, Cecconi2005, Chen2010}. 
Specifically, the SAM-DoA algorithm, designed for the SEAMS mission, enhances DoA below 16 MHz, using a snapshot averaging method and detecting various polarisations \citep{Tanti2023}.
Practical validation experiments indicate that analytically derived DoA algorithms exhibit superior noise immunity and yield satisfactory simulation results. 
However, their effectiveness is constrained by instrument limitations, multi-path interference, and terrestrial signals, introducing non-white noise to the signal \citep{harsha2021ieee, Tanti2023}.
In addition to radio astronomy, DoA estimation has several other applications, such as navigation, tracking, positioning, and search \& rescue operations.

To limit the sensors for the DoA estimation it is crucial to model the signal based on the antenna/ sensor orientation.
Furthermore, single-payload missions in radio astronomy such as Cassini-RPWS show the importance of source localization in order to understand the physical phenomena \citep{Lamy2010}.
This utilises analytical inversion method for DoA estimation.
Similarly in order to reduce the estimation errors methods like MPM-DoA and SAM-DoA where developed for triaxial antenna configuration \citep{Chen2010, Tanti2023}.
However, increasing the estimation sensitivity is observed to result in higher computational complexity and longer inference times.
Also, it is observed that in practical implementations, the signal is distorted due to various environmental or instrument noises \citep{Lamy2010,Tanti2023,Mylonakis2024}.
The induced noise in the signal is mostly non-deterministic, thereby making its analytical modeling difficult.
Acknowledging these issues, ML as a solution is being explored, allowing computers to learn autonomously from data and offering improved accuracy with low inference time compared to some analytical modelled DoA methods.
The idea for such an approach can be traced back to the 1990s \citep{Jha1991, Southall1995, Zooghby2000} and is being researched to date \citep{Liu2018, Wu2019, Kase2020, Wang2024, Mylonakis2024}.
Methods such as DoA estimation using the Hopfield neural network model \cite{Jha1991}, DNN-DoA \cite{Liu2018, Kase2020}, and DCNN for DoA \cite{Mylonakis2024} demonstrate the feasibility and versatility of learning-based algorithms for DoA estimation.
They provide accuracy comparable and even higher than that of analytical models.
However, the developed machine learning techniques for DoA are usually formulated for a linear or two-dimensional array of sensors or antennas.
Having a large number of antennas in an array provides many features that can be extracted to train the machine learning model.
This becomes a challenge when there is a nonuniform array configuration and limited number sensor.
This due to increased complexity in signal modeling in case of nonuniform sensor array and challenge in drawing relations in case of sensor limitation.
In context with the single payload space missions development of any DoA estimation technique is challenging due to the limited number of co-located sensors/antenna on-board and its arrangement.

This paper presents an elementary and novel ML approach for DoA, employing a tri-axial antenna using the artificial neural networks (ANN) framework for radio localisation.
Therefore, integrating it into the SEAMS mission antenna configuration context. \citep{Tanti2023}.
In order to train the model, a synthetic signal with different levels of signal-to-noise ratio (SNR between 0 dB to 60 dB) is generated to model the received signal.
When subjected to the trained ANN architecture, it can achieve a prediction accuracy of approximately 91\%.
Furthermore, the trained ANN-based DoA model demonstrates an inference time that is 80 to 400 times faster than that of analytical models \citep{Tanti2023}.
Thereby, depicting the capability of ML-based approach in the estimation of DoA with a limited number of sensors.

This manuscript is structured as follows. 
Section \ref{method} presents the signal model (Section \ref{sigmod}) and the ANN architecture developed (Sections \ref{annarch} and \ref{ttd}).
Section \ref{rnd} discusses the results of the developed ANN model, and Section \ref{con} provides a conclusion and outlines future work.

\section{Methodology}\label{method}
This section describes the synthetic data generation via an electromagnetic equation incorporating optimal linear antenna characteristics alongside an Artificial Neural Network (ANN) model devised for localization.

\subsection{Received signal model}\label{sigmod}
A substantial volume of data was generated analytically to train our machine learning model.
It includes Fourier components of electric fields, noise, azimuth, elevation, and signal frequency.
The data has been generated by considering a tri-axial linear antenna configuration, positioned orthogonally to one another, considering that the entire configuration is located at the origin.
In the case of a simple linear antenna such as a dipole or monopole, the voltage generated by the antenna due to the received plane wave can be approximated by $V = |\Vec{E}|/h_{\text{ant}}$ where, $h_{\text{ant}}$ is the antenna height, $V$ is the voltage or potential difference across the antenna terminals and electric field ($\Vec{E}$)  of the incident wave \citep{Cecconi2005, balanis2016antenna}.
Thus, considering the antennas of unit length and placed at origin along the orthogonal axes $\hat{a}_x$, $\hat{a}_y$, and $\hat{a}_z$, we can model the electrical field by each antenna due to incident incoherent waves P, traveling towards the origin(reference axes) as,
\begin{equation}\label{eq:e_field}
  E_{m} = \frac{1}{h_{ant}}\sum_{i=1}^{P} \Vec{E_{i}} e^{-j(\Vec{k_i}\cdot\Vec{r_m} - \omega_{i}t)}\cdot\hat{a}_m + n(t)
\end{equation}
Here, ${E_m}$ represents the field due to each antenna element in the tri-axial orthogonal arrangement, $\Vec{E_i}$ is the orientation of the field in three dimension which is perpendicular to the propagation vector $\Vec{k_i}$, $\Vec{r_m}$ is the position vector of each antenna, $\hat{a}_m$ is the unit vector along antenna polarisation that is along the axis of the antenna (where, $m\in(x,y,z)$), and $n(t)$ is the additive white noise due to the media.
Then, the sampled signal is given as,
\begin{equation}\label{eq:e_field_sampled}
  X_m[u] = \sum_{i=1}^{P} \Vec{E_{i}} e^{-j(\Vec{k_i}\cdot\Vec{r_m} - \omega_{i}t_0)} e^{j\omega u\delta}\cdot\hat{a}_m+ n(t_{0}+u\delta)
\end{equation}
Where, the signal is sampled at $t_u = t_0 + u\delta$, $\delta$ being the sampling interval, and $0\leq u\leq N-1$, N is the sample length. 
Thus, to generate the received signal for model training, data has been modelled using equation \ref{eq:e_field} and \ref{eq:e_field_sampled} and by selecting various combinations of azimuth($\phi_{az}$) and elevation($\theta_{el}$) given by the equations \ref{az} and \ref{el}.
\begin{equation}
  \label{az}
  \phi_{az} = 2\pi - tan^{-1}(\frac{k_{y}}{k_{x}}) 
  \end{equation}
\begin{equation}
\label{el}
  \theta_{el} = \pi/2 - sin^{-1}k_{z}
\end{equation}

\subsection{Overview of the ML model}\label{annarch} 
In this paper, we have employed the ANN machine learning architecture due to its versatility to resolve numerical problems.
The initial framework of a neural network consists of three main layers: an input layer, a hidden layer, and an output layer.
The number of hidden layers determines the network's depth, whereas the network's width is determined by the number of neurons in those layers.
An information flow is unidirectional in feed-forward networks as each neuron in one layer is connected to every neuron in the next layer, and the neurons are connected based on weights($w_{ij}$) and bias($b_j$) with $x_j$ being the input data to the jth neuron. 
Analytically, this can also be represented as \citep{Kustrin2000, Russell2016},
\begin{equation}
    Y_i = \sum_{j=1}^{L} w_{ij}x_j + b_j
\end{equation}
The developed ANN architecture comprises nine dense layers, seven of which serve as hidden layers. 
The input layer comprises 16384 neurons, followed by 1024 neurons in the first hidden layer. The remaining layers consist of 512, 64, 32, and 16 neurons, respectively. 
In the output layer, three neurons provide three outputs that indicate the location of the source in azimuth and elevation, along with the frequency at which the emission occurs.
In the developed model, the Rectified Linear Unit (ReLU) is employed as the activation function for each layer.

\subsection{Training and Test Data set}\label{ttd}
\begin{figure}
    \centering
    \includegraphics[width=0.95\linewidth]{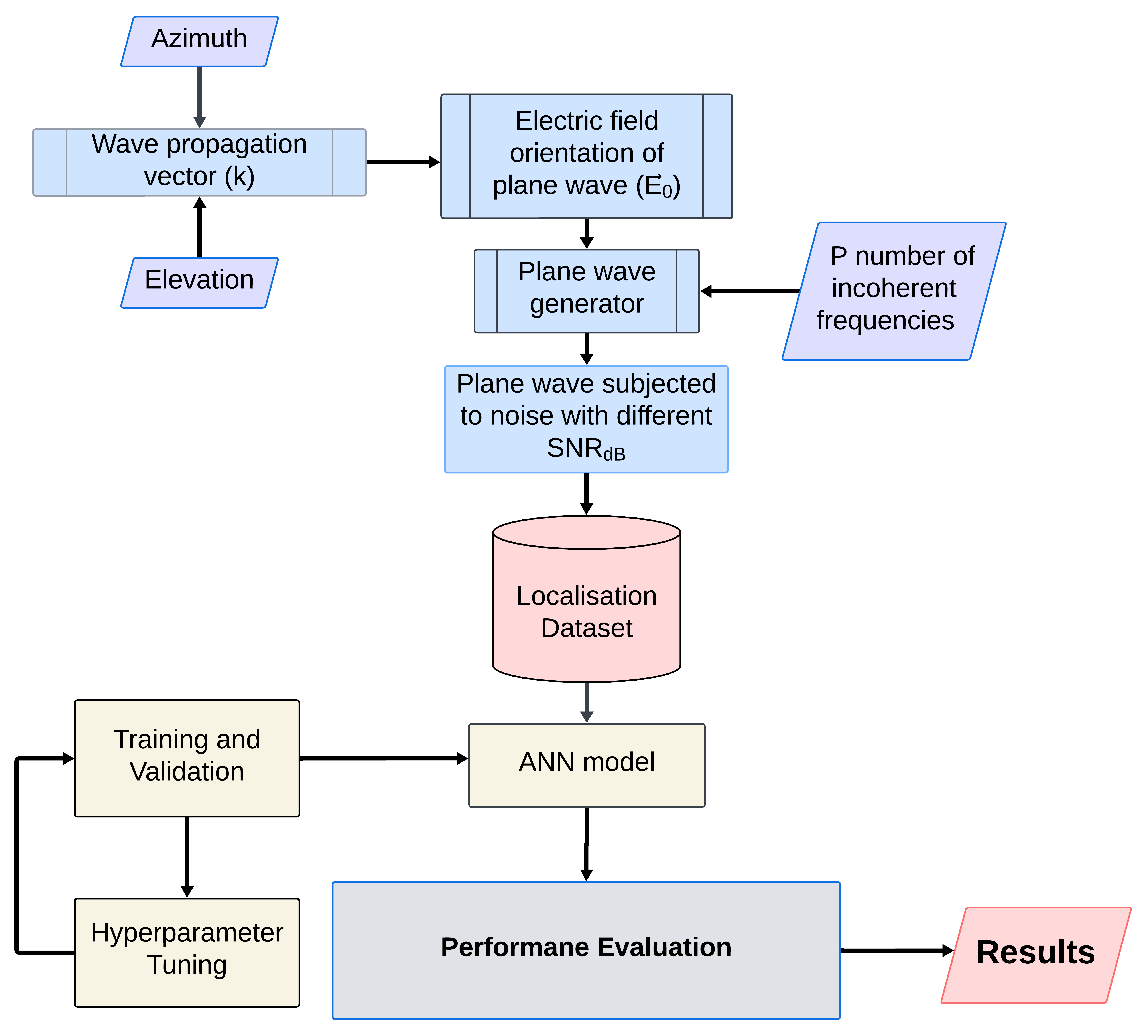}
    \caption{Workflow of the proposed ML model for Radio source localization.}
    \label{fig:flowchart}
    \vspace{-0.1cm}
\end{figure}

The data set for the model's training is generated following the theory in Section \ref{sigmod}. The entire process is depicted in the flow chart shown in Figure \ref{fig:flowchart}.

The data set is created considering that a source is present in the sky that is located anywhere that emits incoherent frequencies in the range of 0.3 to 30MHz. 
Thus to emulate this all plausible combination of azimuth and elevation with a degree-level accuracy is consider for random selection.
This is required in order to train the model to deduce the relation between the acquire signal and DoA angle (analytically related as shown in equation \ref{az} and \ref{el}).
Thereby providing 32,400 DoA angles combination for random selection for signal emulation.
Subsequently, $5\times10^4$ realisation of signals was generated using the method discussed in Section \ref{sigmod} with the DoA angle randomly selected from 32,400 combinations of azimuth and elevation.
Then each realisation was contaminated with noise of 30 different SNR levels between 0 to 60 dB thereby making 30 copies of the same signal with different SNR.
That is the generated plane wave were exposed to noise thereby generating a set of signal having SNR ranging from 0 to 60 dB, as determined by the equation  (\ref{eq:e_field_sampled}), data by each antenna is generated and stored for model training and testing.
There were $1.5\times10^6$ data points to train and test data for different random frequencies and SNRs.
Afterwards, the data was randomly divided into three sets for training, validation, and testing.
The total data was split into 9:1 parts for training and testing, and from the training data set, 10\% was kept aside for model validation.

\begin{table}[!h]
    \centering
    \caption{Test cases for model training with different signal with SNR levels.}
    \begin{tabular}{|p{1cm}|p{2cm}|}
        \hline
        Cases & SNR Range (dB) \\
        \hline
         \hline
         1 & 0 to 60  \\
         \hline
         2 & 20 to 30\\
         \hline
         3 & 30 to 40\\
         \hline
    \end{tabular}
    \label{tab:cases}
\end{table}

The methodology outlined above in Section \ref{annarch} and is the proposed model for DoA estimation using ML that is, Case 1 in Table \ref{tab:cases}. 
In addition to this, two more cases were considered as shown in Table \ref{tab:cases}, wherein the same network was trained for a smaller range of SNR.
In Case 2 and 3, the training data is a subset of data generated for Case 1.
These cases were specifically chosen based on the prediction performance of Case 1 with respect to SNR (Figure \ref{fig:4}) and based on the Flux density spectra of radio emissions from the Earth (Auroral Kilometric Wave (AKR), and Lightning) and the Sun shown in Fig. 3 in \cite{Zarka2012}.
The Fig. 3 in \cite{Zarka2012} clear shows that for the frequency of interest that is 0.3 to 30 MHz the Galactic background will form System Equivalent Flux Density (SEFD) of the system if the receiver noise temperature is maintained below Galactic background noise temperature \citep{Manning2001, Herique2018}.
Thus, the events like AKR, lightning and Solar bursts have SNRs ranging from 0 to 60dB \citep{Zarka2012}.

\section{Results and Discussions}\label{rnd}

In a study of different analytical DoA algorithms, it has been found that accuracy varies with the employed DoA methods.
For example, in case of signals with SNR $\sim$ 15 dB, Snapshot Averaged Matrix Pencil Method (SAM) ($< 0.5^{\circ}$),  Matrix Pencil Method (MPM) ($<1^{\circ}$), and Pseudo-vector estimation-based DoA ($>10^{\circ}$), Analytical Inversion method ($<6^{\circ}$).
Regarding the practical implementation of these algorithms, it has been observed that errors arose to degrees despite the algorithms being capable of achieving precision at the second level \citep{Tanti2023}.
This is caused by the constraints of the instruments and the interference caused by the reflection of signals, which is challenging to represent accurately.
Machine learning provides a feasible alternative by allowing computers to acquire knowledge from data without human intervention. 
On analysing vast data sets, one can identify patterns that enable precise predictions of radio source locations, making it a valuable tool for resolving localisation problems.
 
\begin{figure}[h!]
    \centering
    \includegraphics[width=0.8 \linewidth]{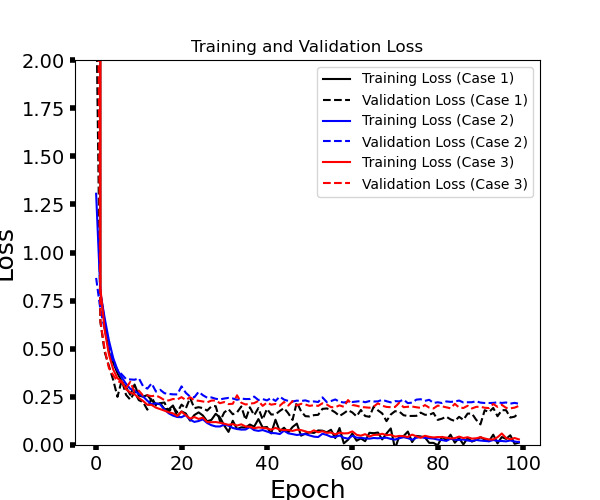}
    \caption{Training and Validation Loss evolution for Different SNR Ranges: The plot displays the training and validation losses over 100 epochs for Case 1 (black, SNR range 0-60 dB), Case 2 (blue, SNR range 20-30 dB), and Case 3 (red, SNR range 30-40 dB). The curves show rapid convergence to low loss values, indicating robust model performance across varying SNR conditions.}
    \label{fig:2}
    \vspace{-0.1cm}
\end{figure}

\begin{figure}[h!]
    \centering
    \includegraphics[width=\linewidth]{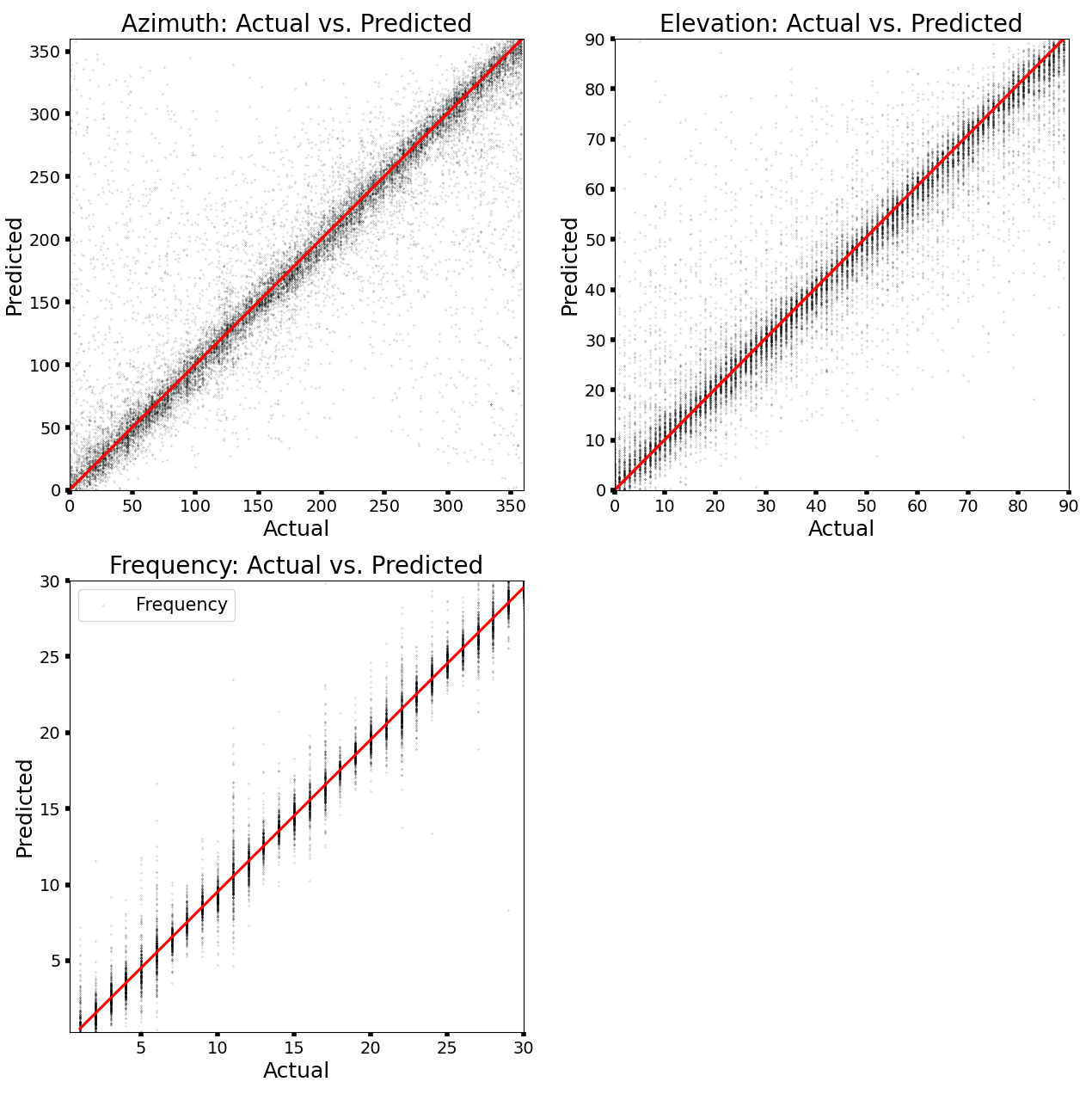}
    \caption{Comparison of Actual to Predicted Values for Azimuth, Elevation, and Frequency: Scatter plots illustrating the relationship between actual and predicted values for Azimuth (top-left), Elevation (top-right), and Frequency (bottom-left). The red line represents the ideal $y=x$ line. The close alignment of points to the red line across all three plots indicates the model's high accuracy in predicting these parameters. The relationship represented here is for Case 1 and it remains similar for the other two cases.}
    \label{fig:3}
    \vspace{-0.5cm}
\end{figure}

In this study, the developed ANN model has been trained for three different cases with a maximum of $1.2\times 10^6$ data points (or signal realisations) for 100 epochs (see Section \ref{ttd}) with the desired output as azimuth, elevation, and signal frequency.
It was observed that the model training starts saturating after 100th epoch thus, the training is halted at 100 epoch, as seen in Figure \ref{fig:2}.
From Figure \ref{fig:2} it is evident that loss of model training and validation converges till $\sim0.02$ \& $\sim0.23$, respectively for Case 1, with an inference time of $\sim5$ ms.
The Table \ref{tab:train_val} shows training and validation loss achieved for all the three test cases.
It can be observed from the Table \ref{tab:train_val} and Figure \ref{fig:2} that the validation loss for Case 1 is consistently lower than the Case 2 and Case 3.
This might be due to the presence of higher SNR signal in the dataset or due to fact that the model is devised for the Case 1.
However, an interesting result is observed when testing for these three test cases is that the if the same model is used to train the signals with different SNR ranges the training and validation loss starts varying. 
In the training and validation evolution figure displays fluctuations.
These drops are mainly due to the smaller batch size in comparison to the data size, optimizer selection, or due to the introduction of random shuffling during training, where both training and validation data sets have been randomly distributed prior to each epoch. 
The purpose of having this shuffling is to ensure impartial model training.

\begin{table}[]
    \centering
    \caption{Training and Validation Loss for three different cases.}
    \begin{tabular}{|p{1.2cm}|p{2.4cm}|p{2.4cm}|}
        \hline
         & Training Loss & Validation Loss\\
         \hline
         \hline
         Case 1 & $\sim0.02$ & $\sim0.23$\\
         \hline
         Case 2 & $\sim0.02$ & $\sim0.26$ \\
         \hline
         Case 3 & $\sim0.02$ & $\sim0.25$\\
         \hline
    \end{tabular} 
    \label{tab:train_val}
\end{table}

Figure \ref{fig:3} represents a scatter plot between the true and predicted azimuth, elevation, and emission frequency values for Case 1 when, the model is subjected to a non-untouched test data set.
It can be observed that the majority of the predictions are around the red line, indicating that the predicted results are equal to actual values.
We had possible combinations of azimuth, elevation and frequency (in the range 0.3 to 30 MHz), causing an almost uniform distribution along the red line in Figure \ref{fig:3}.
Moreover, the merit of any ANN model is quantified using  $R^2$ - value \citep{Madhurima2022}. 
It is a statistical measure of how well the regression line approximates the data.
It is defined by, $R^{2} = 1 - [\sum_{}^{}(y_{i} - \hat{y_{i}})^2]/[\sum_{}^{} (y_{i}-\bar{y})^2]$. 
Here, $y_{i}$ is the actual value, $\hat{y_{i}}$ is the predicted value, and $\bar{y}$ is the mean of the true y values.
Table \ref{tab:r2_tab} shows the $R^2$ of the ANN model for all three cases. 
It is observed that the $R^2$ score remains almost same across the three different cases. 
It is to be note that the $R^2$ score calculate here by inverse scaling the data post prediction hence, the significant drop in value.
Moreover in top left panel of the Figure \ref{fig:3} a increase in scattering is observed towards starting and ending of the Azimuth angle displaying error of 300 degrees and above.
This is due to the fact that the azimuth is an angle and if there is an error of say -20 degree at 0 degree azimuth, it will be represented as 340 degrees. 

\begin{table}[]
    \centering
    \caption{Evaluating $R^2$ score for the trained model during prediction for different cases as shown in Table \ref{tab:cases}.}
    \begin{tabular}{|p{1.8cm}|p{1.5cm}|p{1.5cm}|p{1.5cm}|}
        \hline
        $R^2$ scores & Case 1 & Case 2 & Case 3 \\
        \hline
         \hline
         Azimuth & 0.74 & 0.75 & 0.74 \\
         \hline
         Elevation & 0.76 & 0.78 & 0.79\\
         \hline
         Frequency & 0.96 & 0.97 & 0.98\\
         \hline
    \end{tabular}
    \label{tab:r2_tab}
\end{table}

\begin{figure}[!h]
    \centering
    \includegraphics[width=0.8\linewidth]{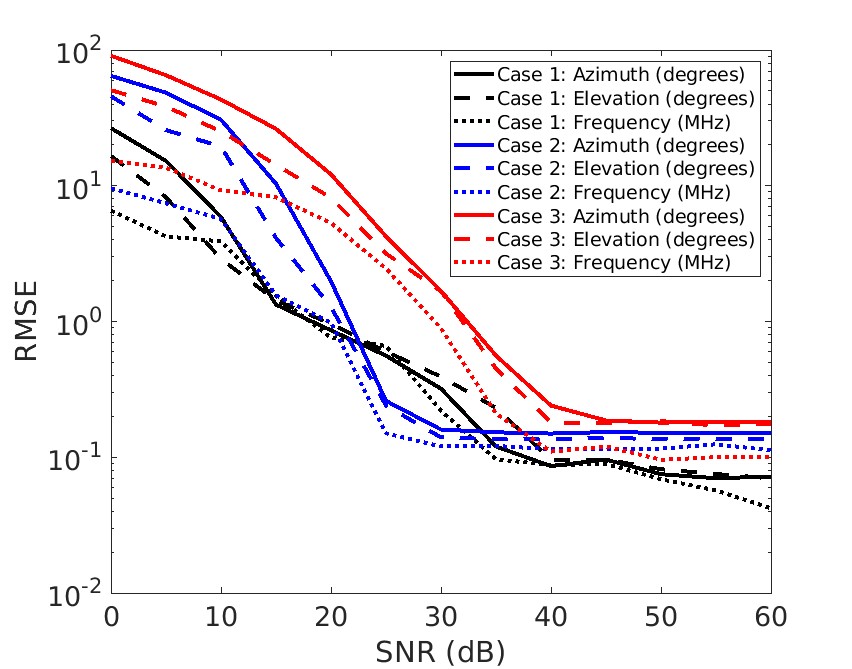}
    \caption{RMSE vs. SNR for Azimuth, Elevation, and Frequency Predictions: This illustrates the RMSE for three different cases as a function of SNR. Case 1 (black), Case 2 (blue), and Case 3 (red) are shown with different line styles representing Azimuth (solid), Elevation (dashed), and Frequency (dotted). The RMSE decreases with increasing SNR, indicating improved prediction accuracy at higher SNR values, with Case 1 showing the best overall performance.}
    \label{fig:4}
\end{figure}

Further, to understand the effect of SNR on the estimation (of azimuth, elevation and frequency of emission), an unseen dataset was segregated into different SNR values and was used to predict the incident signal's azimuth, elevation and frequency.
Figure \ref{fig:4} shows the variation of the estimation error in the Root Mean Squared Error (RMSE) form in the Azimuth, Elevation, and Frequency Predictions.
Figure \ref{fig:4} illustrates the RMSE as a function of SNR for three different cases. 
Each case is shown by a different color (Black for Case 1, Blue for Case 2, and Red for Case 3), while the estimation RMSE in Azimuth, Elevation, and Frequency is indicated by a different line style (solid for Azimuth, dashed for Elevation, and dotted for Frequency).
From Figure \ref{fig:4} it is observed that for Case 2 and 3 the prediction accuracy is very low for lower SNR.
This is expected as the model was retrained for shorter SNR ranges thus its accuracy should be poor for values outside the range.
Interestingly, it was observed that for higher SNR values(that is, for SNR values greater than 30 dB and 40 dB for Case 2 and Case 3 respectively) the estimation RMSE remained nearly constant.
Whereas, the RMSE for the Case 1 is significantly lower than other two cases and decreases consistently till 40 dB thereafter saturating.
Also, it is observed that the model achieves estimation error of less than 1 degree for SNR greater than 15 dB.

\begin{table}[!h]
    \centering
    \caption{Inference time and Error comparison of different DoA Estimation techniques (for Az/El = 180$^\circ$/45$^\circ$, with 4096 number of discrete signal samples, SNR = 15 dB, and clock speed $\sim$2Ghz).}
    \begin{tabular}{|p{3cm}|p{2cm}|p{2cm}|}
        \hline
        Type of DoA Estimator & Inference Time  & Error in Az/El (degrees) \\
        \hline
         \hline
         Elementary ML DoA estimator* & $\sim$6ms & $\sim$1\\
         \hline
         SAM-DoA $(n=20)$ \citep{Tanti2023} & $\sim$6.82m & $\sim$0.015\\
         \hline
         MPM-DoA \citep{Chen2010,Daldorff2009} & $\sim$98s & $\sim$0.0915\\
         \hline
         Analytical Inversion Method \citep{Cecconi2005,Cecconi2007} & $\sim$630ms & $\sim$5.9\\
         \hline
         Pseudo vector estimation based DoA \citep{Carozzi2000} & $\sim$560ms & $\sim$10.2\\
         \hline
    \end{tabular}\\
   \raggedright * The comparison shown is for Case 1 and it does not include the pre and post-scaling time. The pre and post-scaling of the data will add at most $\sim$10ms more to the inference time.
    \label{tab:infer_time}
\end{table}

In a study carried out in \cite{Tanti2023}, it was observed that there are analytical methods with low computational complexity and low accuracy.
It is also observed that to increase the accuracy of such algorithms, the complexity of the method increases exponentially (see \ref{A1}).
Therefore, there is always a trade-off between accuracy and computational complexity. 
However, the ML-based techniques might provide a viable solution as they require high computational costs while training the model. 
In contrast, the trained model has a low computational cost as it assumes the form of multiple linear equations \citep{Agatonovic_kustrin2000, Russell2016}. 
This reduces the computational cost, thereby lowering the inference time.
Table \ref{tab:infer_time} compares the inference time of different analytical algorithms with the developed elementary ML-based DoA estimator.
It is observed that the inference time of this ML-based method is at least $\sim$ ten times faster, including the pre and post-data scaling of the data using a general-purpose computer. 
This inference time will vary if implemented on small form factor machines such as Raspberry Pi, Nvidia Jetson Board, or any other single-board computer (SBC).
In addition, the inference time can be further reduced with the introduction of the new generation of FPGAs capable of hosting ML models.

\section{Conclusion}\label{con}
An elementary novel ML-based DoA model has been proposed for radio source localisation using a tri-axially oriented orthogonal linear antenna.
The ANN architecture has been implemented due to its suitability for handling multi-dimensional simulated data.
The ANN model was trained and tested with $1.2\times 10^6$ data points for 100 epochs (section \ref{ttd}), converging at training \& validating Loss upto $\sim0.02$ \& $\sim0.23$ respectively.
This study exhibits that an ML-based approach is also a viable option for localisation using a finite and limited number of sensors.
Additionally, in the practical implementation of DoA estimation, accuracy, inference time, or detection speed are vital.
The computation time or the inference time of the DoA estimation reaches up to a few seconds \citep{Tanti2023}.
However, using this rudimentary ML-based DoA model, it is observed that the inference time can be reduced to the order of a few milliseconds with higher accuracy (see Table \ref{tab:infer_time}).
The inference time was estimated using a general-purpose computing machine equipped with an Intel Xenon 2nd gen processor. However, this timing will differ across devices; if implemented on a compact machine such as a Raspberry Pi, Nvidia Jetson Board, or another Single Board Computer (SBC), the quoted inference time will be at least ten times longer. Additionally, the inference time can be further decreased with the advent of new-generation FPGAs capable of hosting ML models.
Even though the currently developed model cannot provide polarisation information like SAM-DoA and MPM-based DoA algorithms \citep{Chen2010, Tanti2023}, it is a step towards developing a robust and redundant model.
However, the presented model is an initial prototype; more realistic signal data sets must be modelled or created practically to ensure their accuracy and robustness. 
In addition, it is planned to conduct a more comprehensive study using a better signal model in the future.


\appendix
\section{Computation Speed of SAM-DoA}\label{A1}

\begin{figure}[!h]
    \centering
    \includegraphics[width=0.9\linewidth]{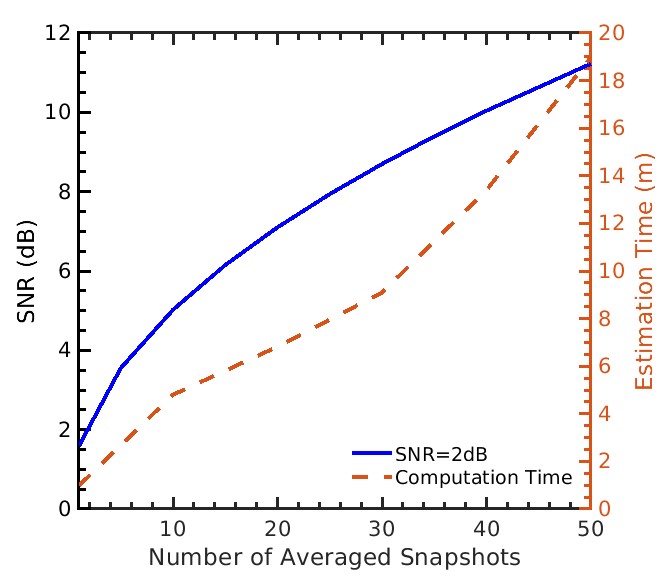}
    \caption{Increase in SNR and Inference time of SAM-DoA algorithm to the number of averaged snapshots.}
    \label{fig:A1-1}
\end{figure}

In \cite{Tanti2023}, an algorithm SAM-DoA is proposed wherein the snapshot averaging method is used to improve the SNR of the signal. 
This addition improved the detection capability in terms of precision of the DoA estimation; however, this came with a cost of increased computational complexity and inference time.
Figure \ref{fig:A1-1} shows how the SNR (solid blue line) and inference / estimation time (orange dotted line) depend on the number of snapshots averaged.
The computational complexity and time of SAM-DoA depend upon the Sample length, the pencil parameter of the matrix pencil method, and the time required to average over snapshots. 
The timing calculations of SAM-DoA in \cite{Tanti2023} were performed for a sample length of 300; however, for practical cases, the sample length is between 1024 and 8192 in multiples of 2. 
For the SEAMS case, the sample length will be 2048 or 4096 based on the Phase I observations. 
In this case, the SAM-DoA will become much slower as for the 4098 sample length, the computation time is around one minute at a clock speed of ~2GHz (number of snapshots averaged = 1). 
If SAM-DoA is to be implemented onboard, wherein the clock speed will be in MHz, the computation time might reach from a few minutes to hours based on the average length.
Thus, even though the SAM-DoA provides a high accuracy when compared to the other algorithms its exponential scaling in inference time with respect to the number of snapshot averaged is a challenge.

\section*{Acknowledgements}
H.A.T. and T.B. thank the entire STARC RF laboratory (IIT Indore) members for their valuable suggestions on the research and manuscript.
A.D. acknowledges the support from the SERB (MTR/2022/000957).
The authors also express their gratitude to the reviewers for their constructive comments.

\bibliographystyle{unsrtnat}
\bibliography{references}  

\end{document}